\documentclass[12pt,epsfig]{article}
\usepackage{amsmath, amssymb}
\usepackage[xdvi]{graphicx}

\makeatletter
\def\tr{\mathop{\rm tr}\nolimits}

\def\diag{\mathop{\rm diag}\nolimits}

\makeatletter

\newcommand{\VEV}[1]{\left\langle #1 \right\rangle}
\newcommand{\del}{\partial}
\newcommand{\drm}{{\rm d}}
\newcommand{\nn}{\nonumber}

\newcommand{\order}[1]{{\cal O}(#1)} 

\newcommand{\TeV}{\mbox{TeV}}

\newcommand{\ie}{{\it i.e.}}

\newcommand{\abs}[1]{\left| #1 \right|}

\newcommand{\bequ}{\begin{equation}}
\newcommand{\eequ}{\end{equation}}
\newcommand{\beqn}{\begin{eqnarray}}
\newcommand{\eeqn}{\end{eqnarray}}
\newcommand{\bctr}{\begin{center}}
\newcommand{\ectr}{\end{center}}
\newcommand{\bit}{\begin{itemize}}
\newcommand{\eit}{\end{itemize}}
\newcommand{\bmtrx}{\begin{pmatrix}}
\newcommand{\emtrx}{\end{pmatrix}}
\newcommand{\Ls}{\left(}
\newcommand{\Rs}{\right)}
\newcommand{\Lm}{\left\{}
\newcommand{\Rm}{\right\}}
\newcommand{\Ll}{\left[}
\newcommand{\Rl}{\right]}
\newcommand{\LL}{\left.}
\newcommand{\RR}{\right.}
\newcommand{\hsp}[1]{\hspace {#1cm}}

\newcommand{\II}{I$\!$I}
\newcommand{\III}{I$\!$I$\!$I}
\newcommand{\IV}{I$\!$V}
\newcommand{\sla}{/\!\!\!}
\begin{document}
\begin{titlepage}
\begin{flushright}
ROMA1/1428/06 \\
SISSA 20/2006/EP \\
hep-ph/0603237
\end{flushright}
\vspace*{1cm}
\begin{center}
{\Large\bf Two-loop Calculation of Higgs Mass\\ 
[0.3cm]
in Gauge-Higgs Unification:} \\
[0.3cm]
{\large\bf 5D Massless QED Compactified on $S^1$}
\end{center}
\vspace{1cm}
\begin{center}
{\large Nobuhito {Maru}}$^{(a),}$
\footnote{E-mail: Nobuhito.Maru@roma1.infn.it}
and
{\large Toshifumi {Yamashita}}$^{(b),}$
\footnote{E-mail: yamasita@sissa.it},
\end{center}
\vspace{0.2cm}
\begin{center}
${}^{(a)}$ {\it Dipartimento di Fisica, 
Universit\`a di Roma "La Sapienza" \\and INFN, Sezione di Roma, 
P.le Aldo Moro 2, I-00185 Roma, Italy}
\\[0.5cm]
${}^{(b)}${\it Scuola Internazionale Superiore 
di Studi Avanzati \\ Via Beirut 4, I-34014 Trieste, Italy}
\end{center}
\vspace{1cm}
\begin{abstract}
We calculate the quantum corrections to the mass of the zero mode of 
the fifth component of the gauge field at two-loop level 
in a five dimensional massless QED compactified on $S^1$. 
We discuss in detail how the divergences are exactly canceled 
and the mass becomes finite. 
The key ingredients to obtain the result are the shift symmetry 
and the Ward-Takahashi identity.
We also evaluate the finite part of corrections. 
\end{abstract}
\end{titlepage}
\newpage
\renewcommand{\theequation}{\thesection.\arabic{equation}} 
\section{Introduction}
Gauge-Higgs unification \cite{FM} is considered to 
be one of the attractive frameworks 
since it provides a solution to the gauge hierarchy problem 
without supersymmetry \cite{gaugehiggs1, gaugehiggs2, SmallVEV, warpgh}. 
In this scenario, the Higgs filed is identified with extra components of 
the gauge field in higher dimensional gauge theories. 
A remarkable feature in the scenario is that quantum corrections to 
the Higgs mass become finite and are independent of the cuttoff scale of 
the theory thanks to the gauge invariance in the higher dimensions 
nevertheless we consider nonrenormalizable theories. 
The Higgs mass is generated through the dynamics of the Wilson line 
for an extra component of the gauge field. 
Noting that the dynamics is nonlocal, we find no counter term 
in the lagrangian, which is assumed to be local, 
to cancel the divergence if the Higgs mass diverges. 
This implies that the Higgs mass should be finite under quantum corrections 
at all order of the perturbations (See Ref. \cite{allorder} for attempts 
to prove the finiteness.). 
Actually, its finiteness at one-loop level was discussed 
by several authors \cite{gaugehiggs1}. 
(In Gravity-Gauge-Higgs unification, the finiteness is guaranteed 
by the general coordinate invariance, see \cite{HLM}.)

Although the concept for the finiteness of the Higgs mass is very clear, 
there are subtleties if we consider higher loop corrections to 
the Higgs mass beyond one-loop level. 
For instance, generally there appear divergences in the subdiagrams 
even if we consider the gauge-Higgs unification scenario. 
These divergences should be subtracted by adding the counter terms 
determined by the lower loop calculations. 
After such a subtraction, the Higgs mass becomes finite 
at any order of perturbations without any additional counter terms. 
This means that the Higgs mass can be predicted even within 
nonrenormalizable theories.
In fact, a Higgs mass at two-loop level are calculated 
in a five dimensional (5D) supersymmetric theory \cite{DGQ}, 
where the linear divergences appear in the one-loop subdiagrams 
and are subtracted by adding one-loop counter terms. 

It is also very important to calculate the Higgs mass beyond one-loop level 
from the phenomenological viewpoint. 
It is known that the physical Higgs mass and the Kaluza-Klein (KK) mass 
tend to be too small in the scenario. 
To get a large KK mass, or in other words to get a small vacuum expectation 
value (VEV) of the Higgs fields compared to the KK mass, 
we rely on a mild tuning to cancel the Higgs mass corrections among 
one-loop contributions \cite{SmallVEV}. 
A large KK mass helps to enhance the physical Higgs mass. 
However, if the KK mass is taken so large, two-loop contributions 
can be important. 
Thus, we can not make the KK mass larger than $\order{10\TeV}$ reliably 
if we do not know the two-loop corrections. 
In this case, the physical Higgs mass can not exceed the present bound 
\cite{LEP} if the low energy effective theory is just the standard model 
\cite{EffTheo}. 
On the other hand, if we control the two-loop corrections, 
the KK mass can be enlarged up to the scale where three-loop 
contributions become important, say $\order{100\TeV}$. 
Then, the physical Higgs mass can pass the experimental test without 
additional low energy degrees. 

As far as we know, there seems no calculation of the Higgs mass 
beyond one-loop order in the context of gauge-Higgs unification. 
Therefore, it is worthwhile to check explicitly the finiteness of 
the Higgs mass for higher order loop corrections. 
In this paper, we explicitly calculate 
the two-loop quantum corrections to the mass 
of the zero mode of the fifth component of the gauge field 
in a 5D massless QED compactified on $S^1$. 
As expected from the general argument of the renormalization theory, 
the mass is shown to be finite. 
A key ingredient to show the finiteness is the shift symmetry 
and Ward-Takahashi identity. 
Although there appear linearly divergent vertex corrections and 
the wave function renormalizations in subdiagrams, 
these divergences are exactly canceled 
as expected from Ward-Takahashi identity. 
In this simple model, 
there is no need to take into account counter terms. 
We will discuss in detail the structure of cancellation of the divergences 
and also evaluate the finite part of the corrections.

This paper is organized as follows. 
In the next section, we introduce our setup and derive Feynman rules. 
Section 3 is the main part of this paper.
Before calculating the two-loop corrections, 
we calculate the one-loop wave function renormalization 
and the vertex corrections to observe that these contributions are 
linearly divergent and have the same magnitude but an opposite sign. 
Then, the two-loop corrections to the mass of the zero mode of 
the fifth component of the gauge field are shown and 
the structure of canceling divergences is clarified. 
The details of this calculation and a physical interpretation 
 are described in Appendix.
The last section is devoted to summarize this paper.

\section{5D Massless QED Compactified on $S^1$}
As an illustration, 
we consider a 5D massless QED compactified on $S^1$ and 
calculate the mass correction to the zero mode of the fifth component 
of the gauge field $A_5$ at two-loop level. 
The action is written as 
\bequ
 S=\int \drm^4 x\drm y \Ll-\frac14 F_{MN}F^{MN} +\bar\Psi i\sla\! D_5\Psi  
                          +{\cal L}_{GF} \Rl,
\eequ
 where $\sla\! D_5=\sla\! D-i\gamma_5D_5$, $\gamma_5^2=1$, 
 $D_M=\del_M-igA_M (M = 0,1,2,3,5)$ is the covariant derivative. 
 $g$ is the 5D gauge coupling constant. 
 We take the mostly minus metric $\eta_{MN}=\diag(+,-,-,-,-)$.
We choose the gauge fixing term as 
\bequ
 {\cal L}_{GF} = -\frac1{2\xi}\Ls \del_\mu A^\mu - \xi \del_5 A^5 \Rs^2,
\label{GaugeFixing}
\eequ
where $\mu = 0,1,2,3$ and $\xi$ is a gauge parameter. 
Then, the gauge part of the action becomes 
\beqn
 S_G&=& \int \drm^4 x \drm y 
       \frac{1}2\Ll -\Ls\del_\mu A_\nu\Rs^2 
             + \Ls1-\xi^{-1}\Rs \Ls\del_\nu A_\nu\Rs^2 
             + \Ls\del_5 A_\nu\Rs^2 \RR \nn\\
    &&\phantom{\int \drm^4 x \drm y\frac{1}2} 
       \LL +\Ls\del_\mu A_5\Rs^2 - \xi \Ls\del_5 A_5\Rs^2 \Rl.
\eeqn
Expanding the gauge field in terms of the Kaluza-Klein modes, 
\beqn
 A_\mu(x^\mu,y) &=& \frac1{\sqrt{L}} \sum_{n=-\infty}^\infty 
                    A_\mu^{(n)}(x^\mu) \exp(2\pi i n\frac{y}{L}), \\
 A_5(x^\mu,y) &=& \frac1{\sqrt{L}} \sum_{n=-\infty}^\infty 
                    A_5^{(n)}(x^\mu) \exp(2\pi i n\frac{y}{L}), 
\eeqn
 where ${A_M^{(n)}}^*=A_M^{(-n)}$ and $L=2\pi R$ is the circumference of 
 the $S^1$, 
 it is written as
\beqn
 S_G&=&\int\drm^4 x \sum_{n=-\infty}^\infty \frac{1}{2}
           \Ll -\abs{\del_\mu A_\nu^{(n)}}^2 
             + \Ls1-\xi^{-1}\Rs \abs{\del_\nu A_\nu^{(n)}}^2 
             + M_n^2\abs{A_\nu^{(n)}}^2 \RR \nn\\
    &&\phantom{\int \drm^4 x \sum_{n=0}^\infty\frac{-1}{2^{\delta_{0n}}}} 
       \LL +\abs{\del_\mu A_5^{(n)}}^2 - \xi M_n^2\abs{A_5^{(n)}}^2 \Rl,   
\eeqn
 where $M_n=2\pi n/L=n/R$ is the KK mass.
\begin{figure}
  \begin{center}
  \includegraphics[width=10.5cm]{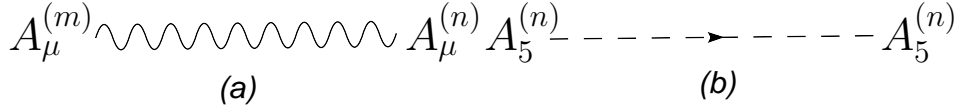}
  \put(-330,25){{\Large $A_\mu^{(m)}$}}
  \put(-180,25){{\Large $A_\mu^{(n)}$}}
  \put(-150,25){{\Large $A_5^{(n)}$}}
  \put(-0,25){{\Large $A_5^{(n)}$}}
 \end{center}
\caption{The propagators of the photon (a) and $A_5$ (b).} 
\label{propagator}
\end{figure}
This leads to the following propagator (see Fig. \ref{propagator}): 
\beqn
(a)    &=& \delta_{mn}
        \Ls \frac{\eta^{\mu\nu}-\frac{p_\mu p_\nu}{M_n^2}}{p^2-M_n^2}
           +\frac{p_\mu p_\nu}{M_n^2}\frac1{p^2-\xi M_n^2}
        \Rs, \label{GaugePropa2}\\
(b)    &=& \frac{-\delta_{mn}}{p^2-\xi M_n^2}. 
\eeqn
Next, expanding the fermion in terms of the KK modes, 
\beqn
 \bar\Psi(x^\mu,y) &=& \frac1{\sqrt{L}} \sum_{n=-\infty}^\infty 
                        \bar\Psi^{(-n)}(x^\mu) \exp(i2\pi n\frac{y}{L}), \\
 \Psi(x^\mu,y) &=& \frac1{\sqrt{L}} \sum_{n=-\infty}^\infty 
                    \Psi^{(n)}(x^\mu) \exp(i2\pi n\frac{y}{L}),
\eeqn
 the fermion part is written as
\bequ
 S_m = \int \drm^4 x \sum_{m,n}
            {\bar\Psi^{(m)}}
            \Ls i\delta_{nm}\Ls \sla\del+M_n\gamma_5\Rs  
                + \sum_{l}\delta_{m\ l+n} 
                 \Ls g_4\sla\! A_\mu^{(l)}-ig_4\gamma_5A_5^{(l)}\Rs
            \Rs\Psi^{(n)}
\eequ
where the 4D gauge coupling constant $g_4$ is defined as $g_4=g/\sqrt{L}$. 
\begin{figure}
 \begin{center}
  \includegraphics[width=14.5cm]{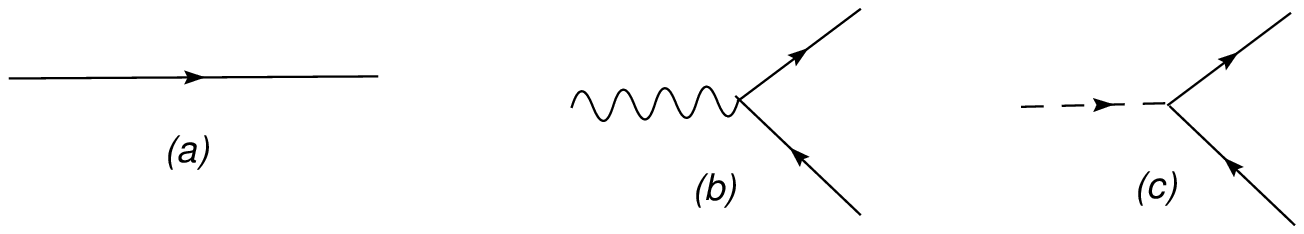}
  \put(-445,45){{\Large $\bar\psi^{(m)}$}}
  \put(-290,45){{\Large $\psi^{(n)}$}}
  \put(-260,35){{\Large $A_\mu^{(l)}$}}
  \put(-135,70){{\Large $\bar\psi^{(m)}$}}
  \put(-135,-5){{\Large $\psi^{(n)}$}}
  \put(-115,35){{\Large $A_5^{(l)}$}}
  \put(5,70){{\Large $\bar\psi^{(m)}$}}
  \put(5,-5){{\Large $\psi^{(n)}$}}
 \end{center}
\caption{Feynman diagrams relevant for the fermion. 
(a), (b) and (c) are the fermion propagator, the gauge interaction vertex 
and the vertex of fermion-fermion-$A_5$, respectively.} 
\label{fermion}
\end{figure}
This leads to the following Feynman rule (see Fig. \ref{fermion}):
\beqn
(a)    &=& \frac{-\delta_{mn}}{\sla p+iM_n\gamma_5} 
      = -\delta_{mn}\frac{\sla p+iM_n\gamma_5}{p^2-M_n^2}, \\
(b)    &=& g_4\delta_{m\ l+n}\gamma_\mu, \\
(c)   &=& -ig_4\delta_{m\ l+n}\gamma_5.
\eeqn
\section{Loop Calculations}
\subsection{One-loop}
Before calculating two-loop corrections, 
we clarify the nature of divergences at one-loop level 
since the divergences appearing in the subdiagrams of two-loop diagrams 
have to be subtracted by adding the counter terms generally. 
The possible relevant counter terms at this order 
 are those of the fermion propagator, 
 the gauge-fermion-fermion vertex and the gauge propagator. 
The first one corresponds to that 
for fermion wave function renormalization. 
The second one corresponds to that for the gauge interaction 
vertex correction. 
The last one should correspond to the renormalization of the 
 gauge coupling. 
\subsubsection{Fermion Wave Function Renormalization}
The wave function renormalization of the fermion is calculated as 
\begin{figure}
 \begin{center}
  \includegraphics[width=14.5cm]{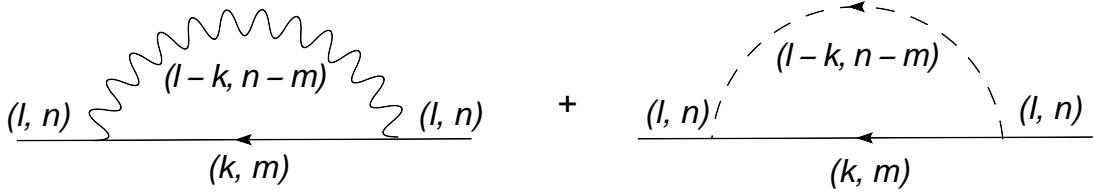}
 \end{center}
\caption{Wave function renormalization of the fermion. 
The corresponding 4D and KK momenta are denoted as $(k, m)$ for example.} 
\label{wfr}
\end{figure}
\beqn
{\rm Fig.}~\ref{wfr} &=& \int\frac{\drm^4 k}{i(2\pi)^4}\sum_{m} g_4^2 
      \Ll
      \gamma_\mu \frac{\sla k+iM_{m}\gamma_5}{k^2-M_{m}^2}
      \gamma_\nu \frac{\eta^{\mu\nu}}{(l-k)^2-M_{n-m}^2}
      \RR\nn\\
  &&\phantom{\int\frac{\drm^4 k}{i(2\pi)^4}\sum_{m} g_4^2}\LL
   + (-i)\gamma_5 \frac{\sla k+iM_{m}\gamma_5}{k^2-M_{m}^2}
     (-i)\gamma_5 \frac{-1}{(l-k)^2-M_{n-m}^2}
    \Rl\nn\\
  &=& \frac{g_4^2}{R}
      \int\frac{\drm^4 k}{i(2\pi)^4}\sum_{m} 
      \frac{-3\Ls\sla k+im\gamma_5\Rs}
           {(k^2-m^2)((l-k)^2-(n-m)^2)},
\label{QEDFermionCorele}
\eeqn
where we normalized all the dimensionful parameters by $1/R$ 
 in the last equation so that all the parameters become dimensionless.
The gauge parameter is taken to be $\xi=1$. 

By using the Feynman integral 
\bequ
 \int_0^1\drm x \left[ \frac{1}{b+(a-b)x} \right]^2
  =\frac1{ab},
\label{FeynmanInt11}
\eequ
 the correction (\ref{QEDFermionCorele}) is written as 
\beqn
 && \frac{g_4^2}{R}
      \int\frac{\drm^4 k}{i(2\pi)^4}\sum_{m} \int_0^1\drm x
      \frac{-3\Ls\sla k+im\gamma_5\Rs}
           {\Ls(k-xl)^2-(m-xn)^2+(x-x^2)(l^2-n^2)\Rs^2} \nn\\
  &=&\frac{g_4^2}{R}
      \int\frac{\drm^4 k'}{i(2\pi)^4}\sum_{m} \int_0^1\drm x
      \frac{-3\Ls x\sla l+im\gamma_5\Rs}
           {\Ls k'^2-(m-xn)^2+(x-x^2)(l^2-n^2)\Rs^2}. 
\label{QEDFermionCoreleFeynman}
\eeqn
Here, we neglect the term that vanishes by the angular integration.

Now we carry out the infinite sum with respect to $m$. 
For this purpose, it is convenient to rewrite the summation 
 by the contour integral in the complex plane, 
\bequ
 \sum_m f(m) \to \int_{C_0}\drm z\frac1{1-\exp(2\pi iz)}f(z)
               = \int_{C_0}\drm z\Ls1+\frac1{\exp(-2\pi iz)-1}\Rs f(z),
\eequ
 where $C_0$ is a contour that encircle the real axis clockwise.
If ${\rm Im}~z \exp(-2\pi\abs{{\rm Im}~z})f(z)$ vanishes 
at $\abs{{\rm Im}~z} \to \infty$
 and $f(z)$ has no poles on the real axis but 
 has poles $\{m_+^i\}$ in the upper half plane 
 and poles $\{m_-^j\}$ in the lower half plane, 
 the contour integral can be expressed by the summation of the residues 
 at each pole and integration on the real axis:
\beqn
&&     \sum_i Res.\Lm\frac{2\pi if(z)}{\exp(-2\pi iz)-1};z=m_+^i\Rm
    +\sum_i Res.\Lm\frac{2\pi if(z)}{1-\exp(2\pi iz)};z=m_-^j\Rm \nn \\
  &&+\int_{-\infty}^\infty \drm zf(z).
\label{Sum2Int}
\eeqn
Note that if $f(z)$ is a real function, each $m_+^i$ has 
 a counter part of $m_-^i={m_+^i}^*$, which means that 
 (\ref{Sum2Int}) can be reduced to 
\bequ
     2 {\rm Re}~\Ll\sum_i Res.\Lm\frac{2\pi if(z)}{\exp(-2\pi iz)-1};
     z=m_+^i\Rm \Rl
    +\int_{-\infty}^\infty \drm zf(z).
\label{Sum2Int2}
\eequ
An important point is that the residues always contain the exponential 
 suppression $\exp(- 2 \pi {\rm Im}~m_+^i)$ for a large ${\rm Im}~m_+^i$, 
 leading to finite contributions. 
Thus, as far as we concern the divergent contributions, 
 it is enough to evaluate the integration on the real axis 
 in (\ref{Sum2Int2}).
%
In other words, 
we can replace the 
 summation with respect to $m$ by the integration on the real axis. 
Then, the correction (\ref{QEDFermionCoreleFeynman}) is written as 
\beqn
&&\frac{g_4^2}{R}
      \int\frac{\drm^4 k'}{i(2\pi)^4}\int_{-\infty}^\infty\drm z_m 
      \int_0^1\drm x
      \frac{-3\Ls x\sla l+iz_m\gamma_5\Rs}
           {\Ls k'^2-(z_m-xn)^2+(x-x^2)(l^2-n^2)\Rs^2} \nn\\
 &=&\frac{g_4^2}{R}
      \int\frac{\drm^4 k'}{i(2\pi)^4}\int_{-\infty}^\infty\drm z_m' 
      \int_0^1\drm x
      \frac{-3\Ls x\sla l+ixn\gamma_5\Rs}
           {\Ls k'^2-z_m'^2+(x-x^2)(l^2-n^2)\Rs^2} \nn.
\eeqn
This shows that the divergent parts of 
 the wave function renormalization and the mass renormalization
 (times $R$) for the fermion mode with $(l,n)$ are commonly given by 
\beqn
  \delta_{Wf}
  &=&{g_4^2}
      \int\frac{\drm^4 k'}{i(2\pi)^4}\int_{-\infty}^\infty\drm z_m' 
      \int_0^1\drm x
      \frac{-3x}
           {\Ls k'^2-z_m'^2+(x-x^2)(l^2-n^2)\Rs^2} \nn\\
  &=&{g_4^2}
      \int\frac{\drm k_E'}{8\pi^2}
      \frac{-3\pi{k_E'}^2}
           {4k_E'^2+l_E^2+n^2} 
\nn\\  &\to&{g_4^2}
       \int\frac{\drm k_E'}{8\pi^2}
       \Ll \frac{-3\pi}4 + {\cal O}(k_E^{-2})\Rl~(k_E' \to \infty),
\eeqn
 where we use the same parameter $k_E'$ for denoting the absolute 
 value of the Wick rotated vector $k_E'$\footnote{
In the next subsection, we use another notation.}.
We find that this correction is linearly divergent.

\subsubsection{Vertex Correction}
The correction to the gauge-fermion-fermion vertex is calculated as 
\begin{figure}
 \begin{center}
  \includegraphics[width=10.5cm]{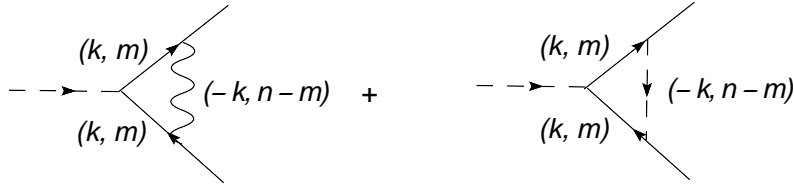}
 \end{center}
\caption{Vertex correction. 
The corresponding 4D and KK momenta are denoted as $(k,m)$ 
for example.} 
\label{vc}
\end{figure}
%
\beqn
{\rm Fig.}~\ref{vc} &=& g_4^2\int\frac{\drm^4 k}{i(2\pi)^4}\sum_{m}
      \Ll
      \gamma_\mu \frac{\sla k+i{m}\gamma_5}{k^2-{m}^2}
      (-i)\gamma_5 \frac{\sla k+i{m}\gamma_5}{k^2-{m}^2}
      \gamma_\nu\frac{\eta^{\mu\nu}}{k^2-({n-m})^2}
      \RR\nn\\
  &&\phantom{\int\frac{\drm^4 k}{i(2\pi)^4}\sum_{m} g_4^2}\LL
   +  (-i)\gamma_5 \frac{\sla k+i{m}\gamma_5}{k^2-{m}^2}
      (-i)\gamma_5 \frac{\sla k+i{m}\gamma_5}{k^2-{m}^2}
      (-i)\gamma_5 \frac{-1}{k^2-({n-m})^2}
    \Rl \nn\\
  &=& {g_4^2}
      \int\frac{\drm^4 k}{i(2\pi)^4}\sum_{m} 
      (-i)\gamma_5
      \frac{3\Ls k^2+m^2\Rs}
           {(k^2-m^2)^2(k^2-(n-m)^2)}
\label{QEDFermionVertex}
\eeqn
where we take the momenta of external lines to be zero. 

Now concentrating on the divergence, we replace the summation 
 with respect to $m$ by the integration on the real axis. 
By carrying out the Wick rotation and 
 using the Feynman integral
\bequ
 \int_0^1 \drm x \frac{2!(1-x)}{((1-x)a+xb))^3}
 = \frac1{a^2b},
\label{FeynmanInt21}
\eequ
 the correction to the vertex $\delta_V$ becomes 
\beqn
 \delta_V
  &=& {g_4^2}
      \int\frac{\drm k_Ek_E^3}{8\pi^2}\int_{-\infty}^\infty\drm z_m'
      \int_0^1\drm x
      \frac{3\Ls k_E^2-(z_m'+xn)^2\Rs\times2!(1-x)}
           {(k_E^2+{z_m'}^2+(x-x^2)n^2)^3} \nn\\ 
  &=& {g_4^2}
      \int\frac{\drm k_E}{8\pi^2}
      \frac{3\pi k_E^2\Ls 4k_E^2-n^2\Rs}
           {(4k_E^2+n^2)^2}  
\nn\\
  &\to& {g_4^2}
      \int\frac{\drm k_E}{8\pi^2}
       \Ll \frac{3\pi}4 + {\cal O}(k_E^{-2})\Rl~(k_E \to \infty). 
\eeqn
We find that it is linearly divergent and is 
 the same as the minus of that of $\delta_{Wf}$, as expected 
 from Ward-Takahashi identity.
This fact is very important to cancel divergences appearing 
in the subdiagrams, as will be seen in the next subsection. 

\subsubsection{Gauge Self Energy}
\begin{figure}
 \begin{center}
  \includegraphics[width=10.5cm]{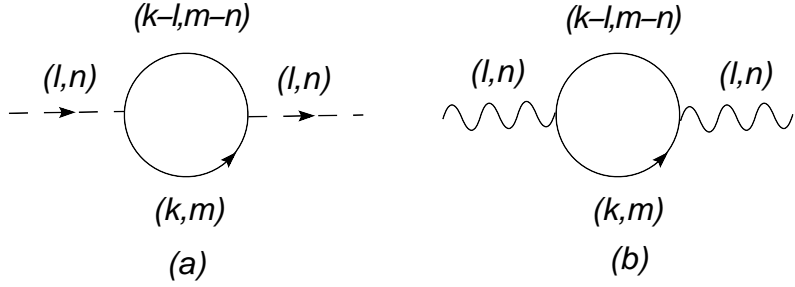}
 \end{center}
\caption{One-loop renormalizations for two-point function 
of $A_5$ (a) and the photon (b).} 
\label{1-loop}
\end{figure}
In this subsubsection, 
we calculate the wave function renormalizations of $A_5$ and $A_\mu$. 
If we denote them as $Z_5$ and $Z_\mu$, respectively, 
these can be expressed at one-loop level symbolically,
\beqn
Z_5 &=& 1 + g_4^2(\Lambda + c) + {\cal O}(g_4^4), 
\label{z5} \\
Z_\mu &=& 1+ g_4^2(\Lambda + c') + {\cal O}(g_4^4)
\label{zmu}
\eeqn
where $\Lambda$ is a cutoff scale of the theory. 
These factors are linearly divergent.  
$c$ and $c'$ mean 
the physical renormalization factors after subtracted the divergence. 
Taking into account these renormalizations, 
the physical Higgs mass at two-loop level includes
\beqn
m^2_{phys@2-loop} &=& 
\frac{g_4^2}{Z_5}m^2_{H@1-loop} + \frac{g_4^4}{Z_5}m^2_{H@2-loop} 
    + {\cal O}(g_4^6) \nn\\
 &=& \frac{Z_\mu}{Z_5}g_R^2 m^2_{H@1-loop} 
    + g_R^4m^2_{H@2-loop} + {\cal O}(g_R^6) 
\nonumber \\
&=& \left[ 1 + g_R^2(c-c') \right] g_R^2 m^2_{H@1-loop}
      + g_R^4m^2_{H@2-loop} + {\cal O}(g_R^6) 
\label{finite}
\eeqn
where the renormalized gauge coupling $g_R$ is defined as 
$g_4^2=g_R^2 Z_\mu$. 
$m^2_{H@1-loop}$ is a one-loop finite mass of the zero mode of $A_5$ 
arising from the diagram in Fig. \ref{1-loop} (a) with zero external momentum. 
Here we define such that $m^2_{H@1-loop}$ does not include the gauge coupling. 
$m^2_{H@2-loop}$ is a two-loop mass which we will evaluate in the next 
 subsection.
Note that the ultraviolet (UV) divergences appearing in (\ref{z5}) 
and (\ref{zmu}) are guaranteed to be the same 
by the five dimensional Lorentz invariance. 
Below, we show it concretely. 
In addition, we will obtain, apart from $m^2_{H@2-loop}$, a finite mass of 
the zero mode of $A_5$ at two-loop level 
which is proportional to $m^2_{H@1-loop}$ and to the difference 
of the finite part, $c-c'$.
Thus, we would like to evaluate also the finite parts of $Z_\mu$ and $Z_5$ and 
 $m^2_{H@1-loop}$, not only the divergent part.
However, note that this contribution should be discriminated 
 from $m^2_{H@2-loop}$.
This is because this contribution does not 
 modify essentially the structure of the one-loop effective potential 
 which is 
 written in terms of $\cos(q g A_5)$, reflecting the phase structure of 
 the Wilson line, where $q$ is a constant.
In other words, the effect merely 
 scales the effective potential in the horizontal direction
 and it is understood by replacing $g_R$ in the potential 
 by $g_R^H=g_R\sqrt{Z_\mu/Z_5}$.

The wave function renormalizations of $A_5$ and $A_\mu$ are shown 
in Fig. \ref{1-loop} (a) and (b), and are calculated as
\beqn
(a) &=& (-1)\int\frac{d^4k}{(2\pi)^4 i}\sum_m 
{\rm tr}\left[
(-i g_4 \gamma_5)\frac{-(k \!\!\!/ + iM_m \gamma_5)}{k^2-M_m^2}
(-i g_4 \gamma_5)\frac{-(k \!\!\!/ - l \!\!\!/ 
+ iM_{m-n} \gamma_5)}{(k-l)^2-M_{m-n}^2}
\right] \nonumber \\
&=& -\frac{ 4 g_4^2}{R^2} \int \frac{ d^4 k_E'}{(2\pi)^4} \sum_m 
\int_0^1 dx \frac{-{k_E'}^2 - x(x-1){l_E}^2 + m^2}
{\Ls{k_E'}^2 + m^2 + (x-x^2){l_E}^2\Rs^2} 
\label{a52p} 
\eeqn
and 
\beqn
(b) &=& (-1)\int\frac{d^4k}{(2\pi)^4 i}\sum_m 
{\rm tr}\left[
(g_4 \gamma_\mu)\frac{-(k \!\!\!/ + iM_m \gamma_5)}{k^2-M_m^2}
(g_4 \gamma_\nu)\frac{-(k \!\!\!/ - l \!\!\!/ 
+ iM_{m-n} \gamma_5)}{(k-l)^2-M_{m-n}^2}
\right] \nonumber \\
&=& -\frac{4g_4^2}{R^2} \int\frac{d^4k_E'}{(2\pi)^4} \sum_m 
\int_0^1 dx \frac{N_{\mu\nu}} 
{\Ls{k_E'}^2 + m^2 + (x-x^2){l_E}^2\Rs^2}, 
\label{ph2p}
\eeqn
respectively, 
where 
\bequ
N_{\mu\nu} = -2 {k_E'}_\mu {k_E'}_\nu 
  + 2x(1-x) {l_E}_\mu {l_E}_\nu \nonumber 
  +g_{\mu\nu} [{k_E'}^2 + m^2 - x(1-x){l_E}^2].  
\eequ
Here, we performed Wick rotation, omitted the terms that vanish after 
the angular integration of $k_E'$ and put $n=0$ since we are interested 
in the wave function of the zero modes.
In the following, we consider only the term proportional to 
 ${l_E}_\mu {l_E}_\nu$ and set $l_E^2=0$ to evaluate $Z_\mu$.

Let us show the divergent parts of (\ref{a52p}) and (\ref{ph2p}), 
which are evaluated by replacing the summation to the integral as before, 
are the same.
Carrying out the integration, we find
\beqn
\int_0^\infty dz_m 
\frac{-{k'_E}^2 - x(x-1)l_E^2 + z_m^2}{({k'_E}^2+z_m^2+x(1-x)l_E^2)^2}
  &=& -\pi \frac{x(1-x)l_E^2}{({k'_E}^2+x(1-x)l_E^2)^{3/2}}, 
\label{a5div} \\
\int_0^\infty dz_m 
\frac{2x(1-x)} {({k_E'}^2 + m^2)^2}
  &=& -\pi \frac{x(1-x)}{{k'_E}^3}. 
\label{phdiv}
\eeqn
From (\ref{a5div}), we can see that this part does not contribute 
 $m^2_{H@1-loop}$, 
 and the contribution to the wave function renormalization is
\bequ
  -\pi \frac{x(1-x)}{{k'_E}^3} 
\eequ
which, including the finite part, is exactly same as (\ref{phdiv}). 
Note that the integration over $z_m$ 
 corresponds to the calculation in the case where the fifth momentum is 
 continuous, \ie\ the limit $R\to\infty$.
In this decompactification limit, the 5D Lorentz symmetry, which is 
 softly broken by the compactification, recovers.
Therefore, 
the cancellation among these contribution is natural.

Next, we evaluate the residue parts which are free from 
 UV divergences.
As for the $Z_5$, we get
\beqn
&&  2 {\rm Re}\Ll Res.\Lm 
                 \frac{2\pi i}{\exp(-2\pi i z_m)-1}
                 \frac{-{k'_E}^2 - x(x-1)l_E^2 + z_m^2}
                 {({k'_E}^2+z_m^2+x(1-x)l_E^2)^2}
                 ; z_m=i\sqrt{{k'_E}^2+x(1-x)l_E^2}\Rm\Rl  \nn\\
&=& -\pi \left[ \frac{ -2x(1-x)l_E^2 }
                     {({k'_E}^2+x(1-x)l_E^2)^{3/2}
                      (e^{2\pi\sqrt{{k'_E}^2+x(1-x)l_E^2}}-1)^2}
         \right. 
\nonumber \\
&& \hsp{0.5}\left. +\frac{4 \pi {k'_E}^2 e^{2\pi \sqrt{{k'_E}^2+x(1-x)l_E^2}}}
{({k'_E}^2+x(1-x)l_E^2)(e^{2\pi\sqrt{{k'_E}^2+x(1-x)l_E^2}}-1)^2} \right]. 
\label{a5conv}
\eeqn
We can find the one-loop correction $m^2_{H@1-loop}$ by setting $l_E^2=0$ 
as, 
\bequ
g_4^2m^2_{A_5@1-loop} = -\frac{4g_4^2}{R^2} \int\frac{d^4k'_E}{(2\pi)^4}
\frac{-2\pi^2}{(-1+\cosh(2\pi k'_E))} = \frac{3g_4^2}{4\pi^4 R^2}\zeta(3). 
\eequ
The wave function renormalization comes from the $l_E^2$ term, 
therefore we obtain by differentiating (\ref{a5conv}) 
with respect to $l_E^2$ 
and setting $l_E^2=0$,
\beqn
&&- 4 g_4^2 \int \frac{ d^4 k'_E}{(2\pi)^4} 
\int_0^1 dx 
(-\pi) 
\left[ 
\frac{2x(1-x)}{{k'_E}^3(e^{2\pi k'_E}-1)} 
+ \frac{4 \pi^2 x(1-x)e^{2\pi k'_E}}{{k'_E} (e^{2\pi k'_E}-1)^2} 
\right. \nonumber \\
&& \hsp{4.8}
  \left. - \frac{4\pi x(1-x)e^{2 \pi k'_E}}{{k'_E}^2 (e^{2\pi k'_E}-1)^2} 
+\frac{8 \pi^2 x(1-x)e^{4\pi k'_E}}{k'_E (e^{2\pi k'_E}-1)^3}
\right]. 
\label{wfa5conv}
\eeqn
The overall factor $1/R^2$ disappears on the dimensional grounds 
in the differentiation. 
%
%
The contribution to $Z_\mu$ is calculated as
\beqn
&&  2 {\rm Re}\Ll Res.\Lm 
                 \frac{2\pi i}{\exp(-2\pi i z_m)-1}
                 \frac{2x(1-x)} {({k_E'}^2 + z_m^2)^2}
                 ; z_m=i k'_E\Rm\Rl  \nn\\
&=& 
-4g_4^2 \int\frac{d^4k'_E}{(2\pi)^4} \int_0^1 dx(-\pi)
\left[
\frac{ 2x(1-x)}{{k'_E}^3(e^{2\pi k'_E}-1)} 
-\frac{4 \pi x(1-x) e^{2 \pi k'_E }  }
{ {k'_E}^2 (e^{ 2 \pi k'_E} -1)^2 } 
\right]. 
\label{cunvam}
\eeqn
Note that these terms have the same for as the first term and 
the third term in (\ref{wfa5conv}). 

From these results, we can obtain ${Z_\mu}/{Z_5}$ at one-loop level as
\beqn
\left[\frac{Z_\mu}{Z_5} \right]_{{\rm finite}} 
&=&1- 4 g_R^2 \int \frac{ d^4 k'_E}{(2\pi)^4} 
\int_0^1 dx 
(-\pi) 
\left[ 
 \frac{4 \pi^2 x(1-x)e^{2\pi k'_E}}{{k'_E} (e^{2\pi k'_E}-1)^2} 
+ \frac{8 \pi^2 x(1-x)e^{4\pi k'_E}}{k'_E (e^{2\pi k'_E}-1)^3}
\right] \nn\\ &&+{\cal O}(g_R^4), 
\eeqn
which is in fact free from UV divergences but contains infrared (IR) 
 divergences. 
This is because we consider exactly massless charged fermion for simplicity.
However, we usually consider the case where $A_5$ which is identified as 
the Higgs field get non-vanishing VEV in the gauge-Higgs unification scenario. 
Then, the charged fermions acquires non-vanishing mass, and the IR divergences 
disappear. 
Thus, we recalculate ${Z_\mu}/{Z_5}$ under the non-trivial background, 
$\VEV{A_5}=a/(gR)$, leading to
\beqn
\left[\frac{Z_\mu}{Z_5} \right]_{{\rm finite}} 
&=&1-4g_R^2 \int\frac{d^4 k'_E}{(2\pi)^4}
\frac{\pi^3 \sinh(2\pi k'_E)(\cos^2(2 \pi a) + \cos(2 \pi a) \cosh(2\pi k'_E)-2)}
{3k'_E (\cosh(2\pi k'_E)-\cos(2\pi a))^3} \nonumber \\
&&+{\cal O}(g_R^4)
\nonumber \\
&=& 1-\frac{g_R^2}{12}\ln(2 \pi a) +{\cal O}(g_R^4), 
\label{wfdif}
\eeqn
in the limit $a\to0$.
\subsection{Two-loop}
In this subsection, 
we calculate two-loop corrections to the mass of the 
 zero mode of $A_5$. 
In 5D massless QED, all the divergences at one-loop level 
  are expected to cancel out.
In fact, we have seen explicitly in the previous subsection 
that the divergences 
from the wave function renormalization and the vertex correction are 
exactly canceled as expected from Ward-Takahashi identity. 
Hence, we calculate two-loop diagrams without any counter terms. 
\begin{figure}
 \begin{center}
  \includegraphics[width=14.5cm]{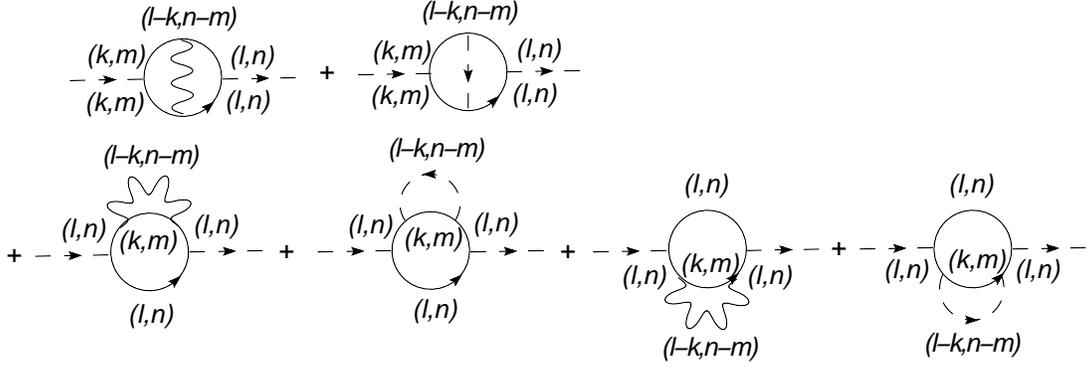}
 \end{center}
\caption{Two-loop diagrams for the mass of the zero mode of $A_5$. 
The corresponding 4D and KK momenta are denoted as $(k,m)$ for example.} 
\label{2loop}
\end{figure}
Straightforward calculation of Fig. \ref{2loop} is given by
\beqn
&&\frac{g_4^4}{R^4}
      \int\frac{\drm^4 l\drm^4 k}{-(2\pi)^8}\sum_{n,m} (-1)\nn\\
  &&\phantom{\int}\Ll
    \tr\Ll (-i)\gamma_5 \frac{\sla l+in\gamma_5}{l^2-n^2}
           \gamma_\mu \frac{\sla k+i{m}\gamma_5}{k^2-{m}^2}
           (-i)\gamma_5 \frac{\sla k+i{m}\gamma_5}{k^2-{m}^2}
           \gamma_\nu \frac{\sla l+in\gamma_5}{l^2-n^2} \Rl
       \frac{\eta^{\mu\nu}}{(l-k)^2-(n-m)^2}
    \RR \nn\\
  &&\phantom{\int}
   +\tr\Ll \gamma_5 \frac{\sla l+in\gamma_5}{l^2-n^2}
           \gamma_5 \frac{\sla k+i{m}\gamma_5}{k^2-{m}^2}
           \gamma_5 \frac{\sla k+i{m}\gamma_5}{k^2-{m}^2}
           \gamma_5 \frac{\sla l+in\gamma_5}{l^2-n^2} \Rl
       \frac{-1}{(l-k)^2-(n-m)^2}
    \nn\\
  &&\phantom{\int}
   +2\tr\Ll (-i)\gamma_5 \frac{\sla l+in\gamma_5}{l^2-n^2}
           \gamma_\mu \frac{\sla k+i{m}\gamma_5}{k^2-{m}^2}
           \gamma_\nu \frac{\sla l+i{n}\gamma_5}{l^2-n^2}
           (-i)\gamma_5 \frac{\sla l+in\gamma_5}{l^2-n^2} \Rl
       \frac{\eta^{\mu\nu}}{(l-k)^2-(n-m)^2}
    \nn\\
  &&\phantom{\int}\LL
   +2\tr\Ll \gamma_5 \frac{\sla l+in\gamma_5}{l^2-n^2}
           \gamma_5 \frac{\sla k+i{m}\gamma_5}{k^2-{m}^2}
           \gamma_5 \frac{\sla l+i{n}\gamma_5}{l^2-n^2}
           \gamma_5 \frac{\sla l+in\gamma_5}{l^2-n^2} \Rl
       \frac{-1}{(l-k)^2-(n-m)^2}
    \Rl \\ 
\label{QEDMassCorr}\
 &=&-12\frac{g_4^4}{R^2}
       \int\frac{\drm^4 l_E\drm^4 k_E}{(2\pi)^8}\sum_{n,m}  \nn\\
  &&\phantom{-12\frac{g_4^4}{R^2} \int}
    \Ll
    \frac{\Ls(l_E^2-n^2)(-k_E^2+{m}^2)-4n{m} l_Ek_E\Rs}
         {(l_E^2+n^2)^2(k_E^2+{m}^2)^2\Ls(l_E-k_E)^2+(n-m)^2\Rs}
    \RR \nn\\
  &&\phantom{-12\frac{g_4^4}{R^2} \int}\LL
   +2\frac{\Ls(l_E^2-n^2)(k_El_E+n{m})-2n^2k_El_E+2n{m}l_E^2\Rs}
          {(l_E^2+n^2)^3(k_E^2+{m}^2)\Ls(l_E-k_E)^2+(n-m)^2\Rs}\Rl,  
\label{QEDMassCorrBFI}
\eeqn
where we note that the contributions from the last two diagrams 
in Fig.~\ref{2loop} are the same as those from the third 
and the fourth diagrams. 
In the last equation, we carry out the Wick rotation.
Now we perform the summations with respect to $n$ and $m$. 
For this purpose, we replace the summations by the integrations 
 on the real axis and the summations of residues, 
 as was done in the one-loop calculation. 
In other words, we decompose the summations of the function $f(m,n)$ 
 to the following four parts:
\beqn
 \mbox{I}   &:& \int\drm z_n\drm z_m f(z_m,z_n), \nn\\
 \mbox{\II} &:& \int\drm z_n 
                2{\rm Re}\Ll\sum_i Res.
                    \Lm\frac{2\pi i f(z_m,z_n)}
                            {\exp(-2\pi i z_m)-1}\,;\,z_m=m_+^i\Rm\Rl, \nn\\
 \mbox{\III}&:& 2{\rm Re}\Ll\sum_i Res.
                    \Lm\frac{2\pi i}{\exp(-2\pi i z_n)-1}
                       \int\drm z_m f(z_m,z_n)\,;\,z_n=n_+^i\Rm\Rl, \nn\\
 \mbox{\IV} &:& 2{\rm Re}\Ll\sum_i Res.\Lm\frac{2\pi i}{\exp(-2\pi i z_n)-1}
                2{\rm Re}\Ll\sum_i Res.\Lm\frac{2\pi if(z_m,z_n)}
                                          {\exp(-2\pi i z_m)-1}
                  \,;\,z_m=m_+^i\Rm\Rl \RR\RR\nn\\
             &&\LL\LL\phantom{2\Re\Ll\Rl\frac{2\pi i}{\exp(-2)}}
               \,;\,z_n=n_+^i\Rm\Rl. \nn
\eeqn
Note that we always carry out the operation of $m$ 
before doing that of $n$. 

Here we list only the results of calculation of each part 
to clarify the cancellation of divergences. 
Detailed calculations and its physical interpretations are 
described in Appendix. 

\beqn
 \mbox{I}   &:& \int\drm z_n\drm z_m f(z_m,z_n) = 0, \label{I2loop}\\
 \mbox{\II} &:& \int\drm z_n 
                2{\rm Re}\Ll\sum_i Res.
                    \Lm\frac{2\pi i f(z_m,z_n)}
                            {\exp(-2\pi i z_m)-1}\,;\,z_m=m_+^i\Rm\Rl \nn\\
     &&= -\frac{4\pi^3(1+e_k)}{e_k^2l((l+\lambda)^2-k^2)}
  -\frac{4\pi^3(1+e_k)}{e_k^2k((k-\lambda)^2-l^2)}\theta(k-\lambda), 
  \label{II2loop}\\
 \mbox{\III}&:& 2{\rm Re}\Ll\sum_i Res.
                    \Lm\frac{2\pi i}{\exp(-2\pi i z_n)-1}
                       \int\drm z_m f(z_m,z_n)\,;\,z_n=n_+^i\Rm\Rl \nn\\
 &&=
 \frac{4\pi^3(1+e_l)}{e_l^2k((k+\lambda)^2-l^2)}
-\frac{4\pi^3(1+e_{k+\lambda})}{e_{k+\lambda}^2k((k+\lambda)^2-l^2))}, 
\label{III2loop}\\
 \mbox{\IV} &:& 2{\rm Re}\Ll\sum_i Res.\Lm\frac{2\pi i}{\exp(-2\pi i z_n)-1}
                2{\rm Re}\Ll\sum_i Res.\Lm\frac{2\pi if(z_m,z_n)}
                                          {\exp(-2\pi i z_m)-1}
                  \,;\,z_m=m_+^i\Rm\Rl \RR\RR\nn\\
             &&\LL\LL
               \,;\,z_n=n_+^i\Rm\Rl 
    =
               -\frac{16\pi^3(1+e_l)\lambda}
          {e_\lambda e_l^2 
           (k+l+\lambda)(k+l-\lambda)(k-l+\lambda)(k-l-\lambda)},
           \label{partIV1}\nn\\
 &&-\frac{4\pi^3(e_k+e_{k+\lambda}+2e_k e_{k+\lambda})}
          {e_k^2e_{k+\lambda}^2 k((k+\lambda)^2-l^2)}
    -\frac{4\pi^3(1+e_{k+\lambda})}
          {e_\lambda e_{k+\lambda}^2 k((k+\lambda)^2-l^2)},
          \label{partIV2}\nn\\
 &&+\frac{4\pi^3(1+e_k)}
         {e_k^2e_\lambda k((k-\lambda)^2-l^2)}
    +\frac{4\pi^3(1+e_k)}
          {e_k^2 k((k-\lambda)^2-l^2)}\theta(k-\lambda) \label{IV2loop}
\eeqn
where $\theta(x)$ is $0$ for $x<0$ and $1$ for $x>0$.
$e_k \equiv \exp(2\pi k) - 1$ and $\lambda \equiv k - l$.

As expected from the five dimensional gauge invariance, 
the contribution from (\ref{I2loop}) vanishes although 
 each term potentially gives divergent correction. 
The first term in (\ref{II2loop}) is the linearly divergent 
term for $l$ momentum, which originated from the vertex correction. 
This divergence is canceled by the first term in (\ref{III2loop}) 
comes from a wave function renormalization. 
All other remaining terms are finite since they are exponentially 
suppressed with respect to $k$ and $l$ momentum.

Now we sum up all the terms of (\ref{I2loop})-(\ref{IV2loop}). 
Note that we can freely exchange $k$ and $l$ with each other keeping 
$\lambda$ unchanged, which 
is nothing but the rename of the integral variables 
 $(k_E,l_E)\to(l_E,k_E)$. 
By using this freedom, we find that the summation becomes zero. 
This shows the finite part corrections vanish, apart from those due to 
 the wave function renormalization of $A_5$. 
This cancellation seems to be accidental in our simple model 
because there is no clear physical reason to ensure such a cancellation. 
If we consider higher order loop corrections beyond two-loops 
even in 5D massless QED or 
calculate quantum corrections in more general models, 
the finite correction would be remained to be nonzero. 
This point would be clarified if we extend our analysis 
to the non-Abelian case, for example \cite{MTY}.

\section{Summary}
Even in gauge-Higgs unification, 
the Higgs mass diverges beyond one-loop level in general. 
The divergence arises from the subdiagrams and should be subtracted 
by adding lower loop counter terms. 
Then, we can obtain the finite Higgs mass at any order of perturbations 
without introducing any other counter terms. 

In this paper, we have calculated quantum corrections to the mass of 
the zero mode of the gauge field at two-loop order 
in a five dimensional massless QED compactified on $S^1$. 
We have found that no counter terms are needed in this simple model 
and have discussed in detail how the possible divergences 
are canceled. 
The key ingredients to obtain such a cancellation 
are the fifth component of 
the 5D gauge symmetry (shift symmetry),  
and the fact that the (linear) divergences 
from the fermion wave function renormalization and the vertex correction are 
the same magnitude with an opposite sign. 
The latter feature is expected from Ward-Takahashi identity. 

We also evaluated the finite part of corrections. 
We classified such corrections to two type: those come from the wavefunction 
renormalization of $A_5$ and those come from 1PI two-loop diagrams.
The former keeps the structure of the one-loop effective potential essentially 
 unchanged and is obtained from the product of the ratio of the wavefunction 
renormalization factors $Z_\mu/Z_5$ and one-loop finite Higgs mass. 
Although these wave function renormalization factors are linearly divergent, 
5D Lorentz invariance ensures that these have same contributions. 
Therefore, the UV divergences are exactly canceled in $Z_\mu/Z_5$ while 
 IR divergences appear.
This is because we consider exactly massless charged fermion, and 
 we introduce a small VEV of $A_5$ as an IR cutoff.
As for the latter, we found that they cancel out among themselves 
 in our calculation.
This result seems to be accidental in our simple model 
because there is no clear physical reason to obtain such a result. 
If we consider higher order loop corrections beyond two-loops 
even in 5D massless QED or 
calculate quantum corrections in more general models, 
the finite correction would be remained to be nonzero. 

We should note that the finite value itself may not be taken seriously 
because our regularization used in this paper does not have 
{\em 4D gauge invariance}. 
Namely, the photon has a non-vanishing mass at one-loop level. 
However, we would like to emphasize that only the 5D Lorentz symmetry
\footnote{In the case of explicit violation of 5D Lorentz invariance as in 
Ref.\cite{SmallVEV}, 
$Z_\mu/Z_5$ may be no longer finite. 
However, in this case, there exist two counter terms to remove 
 the diveregnces in both $Z_\mu$ and $Z_5$. 
Thus, even in such a case, Higgs mass will be finite 
 since the shift symmetry protects its finiteness. 
} 
 ($Z_\mu/Z_5$), the shift symmetry (Part I) 
and the relation expected from Ward-Takahashi identity (Part \II and \III) 
are important to cancel all possible divergences. 
In fact, the 4D gauge invariance is not so important 
for the finiteness of the  mass of $A_5$ 
since the shift symmetry forbids the mass of $A_5$. 
Our regularization indeed preserves the shift symmetry 
by doing the summation of KK modes 
and the relation expected from Ward-Takahashi identity. 
We can conclude from these observations that 
the finiteness for the mass of $A_5$ is correct 
even in our regularization scheme. 
Off course, it is desirable to calculate the mass 
in a full 5D gauge invariant way to obtain a reliable finite mass. 
This subject is left for a future work.

Our discussion of obtaining the finite Higgs mass at any order 
of perturbations would be generic in any Gauge-Higgs unification models. 
Therefore, it would be very interesting to extend our analysis 
to non-Abelian case not only from the theoretical 
but also from the phenomenological viewpoints. 
This subject will be reported elsewhere \cite{MTY}.


\vspace*{1cm}

\begin{center}
{\bf Acknowledgements}
\end{center}  
N.M. would like to thank N. Sakai for bringing my attention to this subject 
and C.S. Lim for valuable discussions at an early stage of this work. 
He also would like to thank L. Silvestrini for useful comments. 
T.Y. would like to thank K. Izawa for stimulating discussions. 
We would like to thank M. Serone for valuable comments on 
the previous version of the manuscript. 
The work of N.M. is supported by INFN, sezione di Roma. 
The work of T.Y. is supported by SISSA. 
\vspace*{1cm}
\begin{appendix}
\section{Detailed calculation of two-loop corrections to Higgs mass}
In this appendix, the detailed calculations of Higgs mass at two-loop part 
expressed as I-\IV in 3.2. 
and its physical interpretations are described. 

\setcounter{equation}{0}
\subsection{Part I}

First, we evaluate the contribution from the first part, namely 
 the summations are replaced by integrations on the real axis. 
This contribution is expected to correspond to the diagrams where
 both loops do not wind around $S^1$ and thus can be shrinked to a point. 
In general, such diagrams give the strongest divergences. 
However, in our case, the five dimensional gauge invariance 
 will forbid such a contribution.

Before evaluating the contribution, we define a new vector 
 from $l_E$ and $k_E$ as $\lambda_E \equiv k_E - l_E$, and we use the same 
 parameters without the index $E$ 
 to denote the absolute values of the vectors. 
Among these three vectors, we can choose any two vectors as the 
 integral variables. 
Then the integrand of (\ref{QEDMassCorrBFI}) is written as 
\beqn
I(m,n) &\equiv&   \frac{(l^2-n^2)(m^2-k^2)-2n{m}\Ls l^2+k^2-\lambda^2\Rs}
         {(l^2+n^2)^2(k^2+{m}^2)^2\Ls \lambda^2+(n-m)^2\Rs} \nn \\
   &&+\frac{(l^2-3n^2)(l^2+k^2-\lambda^2)+2(3l^2-n^2)nm}
          {(l^2+n^2)^3(k^2+{m}^2)\Ls \lambda^2+(n-m)^2\Rs}
\label{QEDMassCorrBFIIntegrand}.
\eeqn
We can integrate over $z_m$ of 
 (\ref{QEDMassCorrBFIIntegrand}) by adding the integration on the large 
 half-circle in the upper half plane and evaluating the residues at the 
 poles on the plane.
\beqn
I(n) &=&
  \int_{-\infty}^\infty\drm z_m I(z_m,n) 
    = Res.\Lm 2 \pi i I(z_m,n) ;
             z_m=ik,n+i\lambda \Rm \nn\\
 &=&-
    \frac{\Ls k(k+\lambda)^2l^2+\Ls(k-2\lambda)(k+\lambda)^2
         +(k+2\lambda)l^2\Rs n^2 +kn^4\Rs\pi}
         {(l^2+n^2)^2 k \lambda\Ls(k+\lambda)^2+n^2\Rs^2}
    \nn\\
  &&+
    \frac{\Ls(k+\lambda)l^2(k^2+l^2-\lambda^2)-3(k-\lambda)\Ls(k+\lambda)^2
         -l^2\Rs n^2-2kn^4\Rs\pi}
         {(l^2+n^2)^3 k \lambda((k+\lambda)^2+n^2)}
\label{QEDMassCorrBFIIntegrandM}.
\eeqn
In a  similar way, we can further perform the integration over $z_n$ 
 of the above expression to find
\beqn
 && \int_{-\infty}^\infty \drm z_n I(z_n) 
    = Res.\Lm 2 \pi i I(z_n) ;
             z_n=il,i(k+\lambda) \Rm \nn\\
 &=&
    -\frac{\Ls k+l-\lambda\Rs\pi^2}
         {kl \lambda\Ls k+l+\lambda\Rs^2}
   +\frac{\Ls k+l-\lambda\Rs\pi^2}
         {kl \lambda\Ls k+l+\lambda\Rs^2}\quad=0.
\eeqn
As expected from the five dimensional gauge invariance, 
the contribution from this part vanishes although 
 each term potentially gives divergent correction.

\subsection{Part \II}
\begin{figure}
 \begin{center}
  \includegraphics[width=4.5cm]{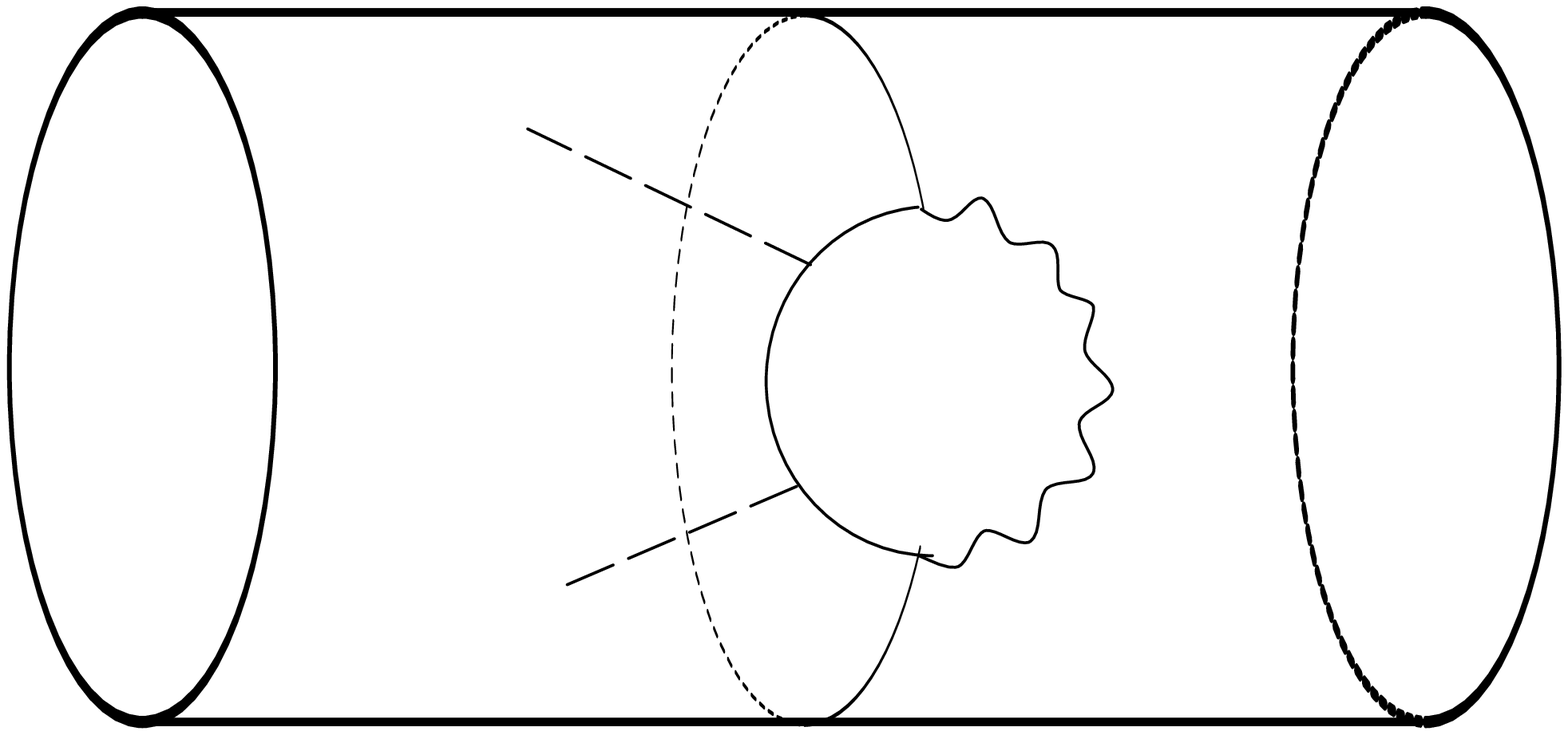}
  \includegraphics[width=4.5cm]{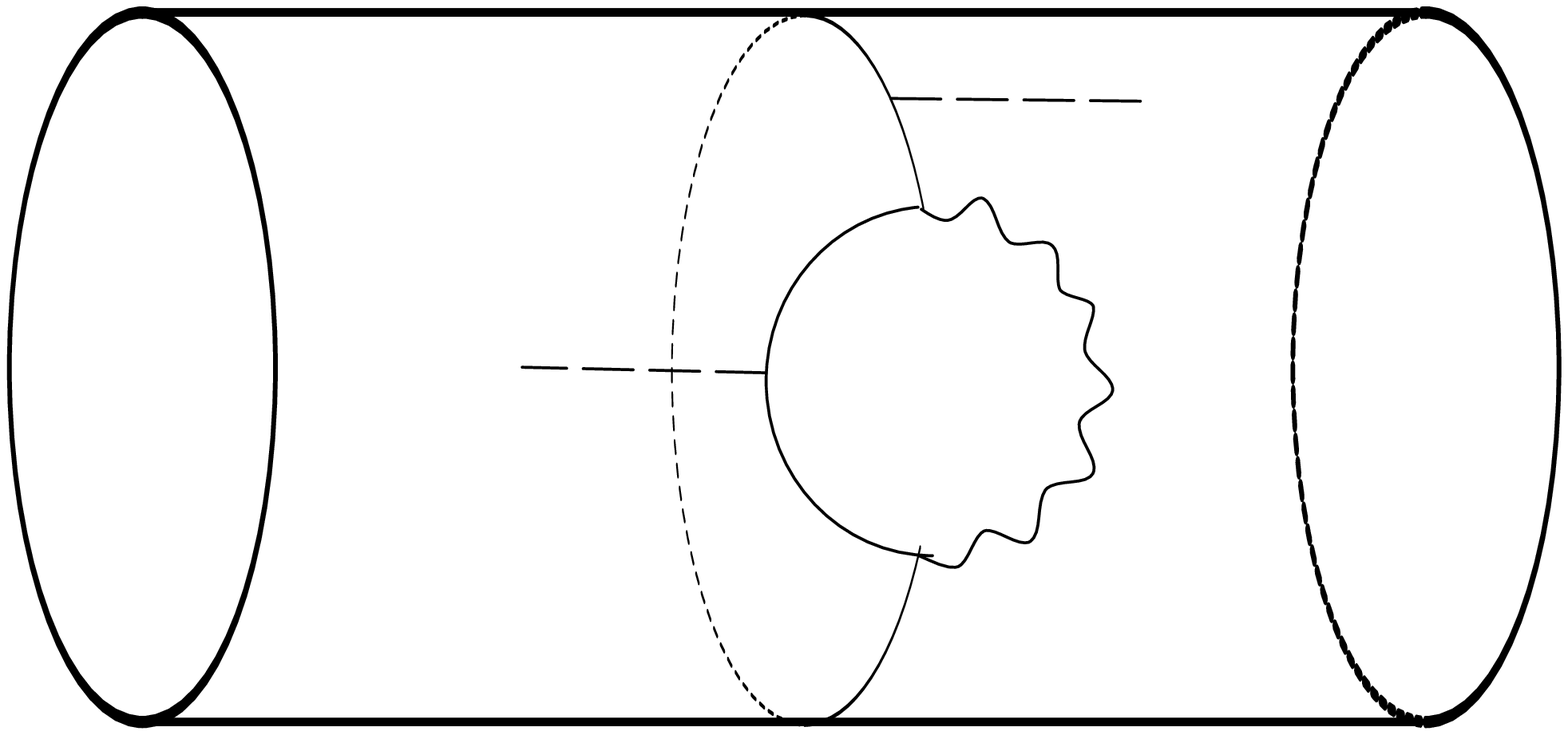}
  \includegraphics[width=4.5cm]{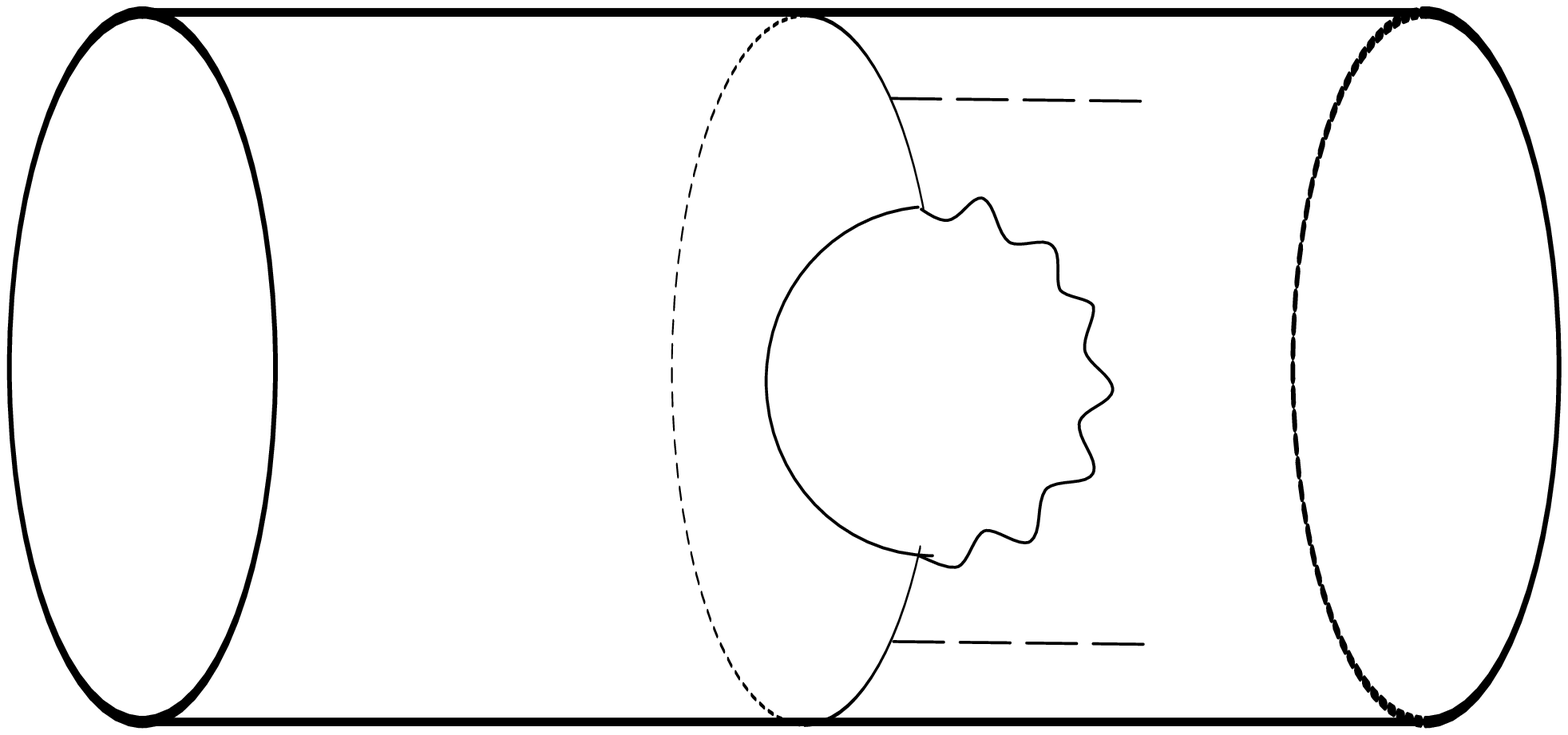}
  \put(-330,5){{\Large (a)}}
  \put(-200,5){{\Large (b)}}
  \put(-70,5){{\Large (c)}}
 \end{center}
 \vspace*{-0.5cm}
\caption{The diagrams where the fermion loop winds around $S^1$ 
but the photon loop does not. 
The cylinder denotes $S^1$. 
If the fermion-photon loop is shrinked to a point, these diagrams provide 
corrections of the 4-point vertex of the fermion-fermion-$A_5$-$A_5$ (a), 
the gauge interaction vertex (b) and the wave function renormalization (c), 
respectively.} 
\label{wrap}
\end{figure}
Next, we evaluate the contribution from the second part.
This contribution is expected to correspond to the diagrams where
 one of the loops winds around $S^1$ while the other does not. 
Some examples are shown in Fig. \ref{wrap}. 
Because the latter loop can be shrinked to a point, 
generally this part gives a divergent contribution, 
even in the gauge-Higgs unification scenario. 
However, as is well known, such divergences can be cancelled by 
 the one-loop counter terms. 
In other words, after we remove all the divergences in the one-loop diagrams, 
 the contribution from this part will be finite. 
Then, we obtain the finite mass at two-loop level without any additional 
counter terms. 
In \cite{DGQ}, the Higgs mass at two-loop level are calculated 
 in 5D supersymmetric theory, where supersymmetry is broken 
 by Scherk-Schwarz mechanism. In fact, the linear divergences appear 
 and are cancelled by the one-loop counter terms. 
In our particular case, 
Ward-Takahashi identity should make 
 the divergence in this contributions same as the minus of 
 the divergence in the contribution from the part \III. 
Thus, all the divergences are expected to cancel out with each other 
 {\em without} any counter terms. 

Now, let us evaluate the residues of the poles of $z_m$ 
 in the upper half plane, \ie\ $z_m=i k$ and $z_m=n+i \lambda$. 
We can interpret the former contribution as the one comes from 
 the diagram where the fermion line with the momentum $k$ winds 
 around $S^1$~(Fig.~\ref{wrap}), 
 while the latter as the one from the diagram where 
 the photon line winds. 
The residue of the first term 
 in (\ref{QEDMassCorrBFIIntegrand}) 
 on the pole $z_m=i k$ is evaluated as 
\beqn
J_1(n) &\equiv& 2{\rm Re}\Ll Res.\Lm\Ls\mbox
                 {1st term of}~I(z_m,n) \Rs
                 \frac{2\pi i}{\exp(-2\pi i z_m)-1}
                 ; z_m=ik\Rm\Rl \nn\\
 &=&  4\pi \left[ e_k\Ls k^2l^2(k^2-\lambda^2)^2
                     +(2k^6+k^4(l^2-3\lambda^2)-4k^2l^2\lambda^2
                      -(l^2-\lambda^2)\lambda^4)n^2 \RR\RR\nn\\
 &&\LL\LL\LL\hsp{1.1}  
                     +(4k^4-k^2(l^2+2\lambda^2)-2(l^2-\lambda^2)\lambda^2)n^4
                     +(2k^2-l^2+\lambda^2)n^6 \Rs
 \RR\RR\nn\\ &&\LL\hsp{0.5}  
             +(1+e_k)k(k^2-\lambda^2+n^2)((k+\lambda)^2+n^2)
                     ((k-\lambda)^2+n^2)(l^2+n^2)\pi \right]   \nn\\
 && /\left[ e_k^2k((k+\lambda)^2+n^2)^2((k-\lambda)^2+n^2)^2(l^2+n^2)^2 
 \right] 
\label{VertexMKwrap}
\eeqn
and that of the second term is given by
\beqn
J_2(n) &\equiv& 2{\rm Re}\Ll Res.\Lm\Ls\mbox
                 {2nd term of}~I(z_m,n) \Rs
                 \frac{2\pi i}{\exp(-2\pi i z_m)-1}
                 ; z_m=ik\Rm\Rl \nn\\
 &=& 2\pi \left[ -(k^2-\lambda^2)^2(k^2+l^2-\lambda^2)^2l^2
  \RR\nn\\&&\LL  
          +(3k^4-2k^2(4l^2+3\lambda^2)+(l^2-\lambda^2)(l^2-3\lambda^2))n^2 
          + (k^2-3l^2+3\lambda^2)n^4 
          \right]
          \nn\\
&& /\left[ e_k k ((k+\lambda)^2+n^2)((k-\lambda)^2+n^2)(l^2+n^2)^3 \right], 
\label{WFMKwrap}
\eeqn
where $e_k\equiv\exp\Ls2\pi k\Rs-1$.
After the integration over $z_n$, we find these respectively become 
\beqn
  &&\frac{2\pi^2\Ls 2(1+e_k)(k+l)((k+l)^2-\lambda^2)\pi
                   +e_k(3(k+l)^2+\lambda^2)\Rs}
         {e_k^2 kl((k+l)^2-\lambda^2)^2}, \\
  &&-\frac{2\pi^2\Ls(3(k+l)^2+\lambda^2)\Rs}
         {e_k kl((k+l)^2-\lambda^2)^2}
\eeqn
 for $k>\lambda$ and
\beqn
  &&\frac{2\pi^2\Ls 2(1+e_k)k\lambda(k^2-(l+\lambda)^2)\pi
                   +e_k(k^2(l+3\lambda)-(l-\lambda)(l+\lambda)^2)\Rs}
         {e_k^2 kl\lambda(k^2-(l+\lambda)^2)^2}, 
\label{VertexMKwrapN}\\
  &&-\frac{2\pi^2\Ls k^2(l+3\lambda)-(l-\lambda)(l+\lambda)^2\Rs}
         {e_k kl\lambda((\lambda + l)^2 - k^2)^2}
\label{WFMKwrapN}
\eeqn
 for $k<\lambda$.
Note that all terms vanish in the limit $e_k\to\infty$. 
This means that we do not have UV divergences in $k$ integration.
On the other hand, we may encounter divergences in $l$($\lambda$) 
 integration. 
In fact, we can see that the integration of (\ref{VertexMKwrapN}) 
 over $l_E$ is linearly divergent, while the one of (\ref{WFMKwrapN}) 
 converges.
These are consistent with the interpretation that these contributions 
 correspond to the diagram where the fermion line with the 
 momentum $k$ winds on $S^1$:
 (\ref{VertexMKwrapN}) corresponds to the vertex correction 
 (Fig.~\ref{wrap} (b)) 
 while (\ref{WFMKwrapN}) corresponds to the correction of the four point 
 vertex fermion-fermion-$A_5$-$A_5$ (Fig.~\ref{wrap} (a)).

In a similar way, contributions from the pole at $z_m=n+i\lambda$ 
 is evaluated as 
\beqn
J_3(n) &\equiv& 2{\rm Re}\Ll Res.\Lm\Ls\mbox
                 {1st term of}~I(z_m,n) \Rs
                 \frac{2\pi i}{\exp(-2\pi i z_m)-1}
                 ; z_m=n+i\lambda\Rm\Rl \nn\\
 &=&- 2\pi\left[ (k^2-\lambda^2)^2(k^2+\lambda^2)l^2
             +((k^2-\lambda^2)^2(k^2+5\lambda^2)
    \RR\nn\\
 &&\LL
             +(3k^4-6k^2\lambda^2-5\lambda^4)l^2)n^2
             +(3k^4+k^2(3l^2+2\lambda^2)-5l^2\lambda^2+3\lambda^4)n^4
\RR \nn\\&&\LL
 +(3k^2+l^2-\lambda^2)n^6 +n^8 \right]
 /\left[ e_\lambda\lambda((k+\lambda)^2+n^2)^2((k-\lambda)^2+n^2)^2(l^2+n^2)^2 
  \right] \nn \\
\label{VertexMKmlwrap}
\eeqn
and that of the second term is given as
\beqn
J_4(n) &\equiv& 2{\rm Re}\Ll Res.\Lm\Ls\mbox
                 {2nd term of}~I(z_m,n) \Rs
                 \frac{2\pi i}{\exp(-2\pi i z_m)-1}
                 ; z_m=n+i\lambda\Rm\Rl \nn\\
 &=& 2\pi\left[ (k^2-\lambda^2)(k^2+l^2-\lambda^2) l^2
             +(-3(k^2-\lambda^2)^2+4(k^2+2\lambda^2)l^2+l^4)n^2
  \RR\nn\\&&\LL  
             +(-5k^2+3l^2+\lambda^2)n^4-2n^6 \right] 
 /\left[ e_\lambda\lambda((k+\lambda)^2+n^2)((k-\lambda)^2+n^2)(l^2+n^2)^3 
 \right], \nn\\
\label{WFMKmlwrap}
\eeqn
where $e_\lambda\equiv\exp\Ls2\pi \lambda\Rs-1$.
After the integration over $z_n$, we find these terms give 
 the same contributions with an opposite sign and the sum of 
 these vanishes.
This is also consistent with the interpretation that these contributions 
 correspond to the diagram where the photon line winds around $S^1$.
Namely, such contributions correspond to the correction of the four point 
 vertex $A_M$-$A_M$-$A_5$-$A_5$. 
This is the correction to the $F_{MN}^4$ term 
which has vanishing contribution to the mass correction 
because its Feynman rule contains momenta of the four lines 
and our interest is zero external momenta case. 

In summary, the contribution from this part is written as 
\bequ
 -\frac{4\pi^3(1+e_k)}{e_k^2l((l+\lambda)^2-k^2)}
 -\frac{4\pi^3(1+e_k)}{e_k^2k((k-\lambda)^2-l^2)}\theta(k-\lambda),
\label{partII}
\eequ
where $\theta(x)$ is $0$ for $x<0$ and $1$ for $x>0$.

The first term is the linearly divergent term for $l$ momentum, 
which originated from the vertex correction. 
This divergence should be canceled 
by the term originated from the wave function renormalization. 
We will see in the next subsection that this is indeed the case. 
On the other hand, the second term is finite 
since this contribution exists only when the momentum $\lambda$ 
is smaller than the momentum $k$. 

\subsection{Part \III}
\label{Part3}
Now, we evaluate the contribution from the third part.
The integration over $z_m$ is given in (\ref{QEDMassCorrBFIIntegrandM}). 
It shows that there are two poles in the upper half plane: 
 $z_n=il$ and $z_n=i(k+\lambda)$.
The contribution from the pole at $z_n=i l$ is calculated as 
\beqn
 && 2{\rm Re}\Ll Res.\Lm I(z_n)
                 \frac{2\pi i}{\exp(-2\pi i z_n)-1}
                 ; z_n=il\Rm\Rl \nn\\
 &=&-\frac{4\pi^3(1+e_l)}
         {e_l^2 k((k+\lambda)^2-l^2)}
   -\frac{2\pi^2\Ls(k-\lambda)(k+\lambda)^2-(k+3\lambda)l^2\Rs}
         {e_l kl\lambda((k+\lambda)^2-l^2)^2}
\label{VertexMNLwrap}\\
  &&+\frac{8\pi^3(1+e_l)}
         {e_l^2 k((k+\lambda)^2-l^2)}
   +\frac{2\pi^2\Ls(k-\lambda)(k+\lambda)^2-(k+3\lambda)l^2\Rs}
         {e_l kl\lambda((k+\lambda)^2-l^2)^2} \nn \\
   &&+\frac{4\pi^4(1+e_l)(2+e_l)(k-\lambda)}
         {e_l^3 kl\lambda}
\label{WFMNLwrap}
\eeqn
 where $e_l\equiv \exp\Ls2\pi l\Rs-1$.
These terms vanish in the limit $e_l\to\infty$, and thus 
 the $l$ integration is free from UV divergences.
Note that if we choose $-l_E$ and $\lambda_E$ as the integral variables 
 and rename them as $L_E$ and $K_E$, 
 $k_E$ is written as $k_E=K_E-L_E\equiv\Lambda_E$. 
Then the three new momenta $(K_E,L_E,\Lambda_E)$ satisfy 
 the same relation as $(k_E,l_E,\lambda_E)$.
In addition, the integration measure under this rename is invariant. 
Thus, this means we can replace $k$ and $\lambda$ with each other. 
From this observation, it is clear that the 4D momentum integral of 
the last term in (\ref{WFMNLwrap}) 
 does not contribute. 

The first terms of (\ref{VertexMNLwrap}) and (\ref{WFMNLwrap}) 
are linearly divergent with respect to $k$ integration.
It is interesting to find that the divergence in (\ref{VertexMNLwrap}) 
 is the half of the one in (\ref{WFMNLwrap}) with the opposite sign, 
 and is the same as the one in the part \II.
These results are again consistent with the interpretation 
 that these contributions correspond to the diagram 
 where the fermion line with the momentum $l$ winds around $S^1$:
 (\ref{VertexMNLwrap}) corresponds to the vertex correction 
 (Fig.~\ref{wrap} (b))
 and (\ref{WFMNLwrap}) corresponds to the wave function correction 
 of the fermion (Fig.~\ref{wrap} (c)). 
 The second terms in (\ref{VertexMNLwrap}) 
 and (\ref{WFMNLwrap}) are canceled. 

The contribution from the pole at $z_n=i(k+\lambda)$ is summarized as 
\beqn
  2 {\rm Re}\Ll Res.\Lm I(z_n) 
                 \frac{2\pi i}{\exp(-2\pi i z_n)-1}
                 ; z_n=i(k+\lambda)\Rm\Rl 
 =  -\frac{4\pi^3(1+e_{k+\lambda})}{e_{k+\lambda}^2k((k+\lambda)^2-l^2))}
 \nn \\
\eeqn
 where $e_{k+\lambda}\equiv\exp\Ls{2\pi(k+\lambda)}\Rs-1$.
This term vanishes when $e_{k+\lambda}\to\infty$, 
 and thus this contribution is finite under both $k$ and $l$ 
 integrations.
This contribution is interpreted as coming from the diagram where 
 the fermion line with the momentum $k$ and the photon line 
 wind around $S^1$.

In summary, the contribution from this part is written as 
\bequ
 \frac{4\pi^3(1+e_l)}{e_l^2k((k+\lambda)^2-l^2)}
-\frac{4\pi^3(1+e_{k+\lambda})}{e_{k+\lambda}^2k((k+\lambda)^2-l^2))}.
\label{partIII}
\eequ
The first term is linearly divergent, which originated from 
two wave function renormalizations and a vertex correction, 
namely a wave function renormalization. 
One can see that this contribution and the first term in (\ref{partII}) 
are exactly canceled as expected from Ward-Takahashi identity.  

\subsection{Part \IV}
Finally we evaluate the contribution from the fourth part.
The contribution of the part of the residues in $m$ is written 
 in (\ref{VertexMKwrap}), (\ref{WFMKwrap}), (\ref{VertexMKmlwrap}) 
 and (\ref{WFMKmlwrap}). 
They have poles on the upper half plane at 
 $z_n=il,i(k+\lambda),i\abs{k-\lambda}$. 
The first and the last two parts give the following contributions;
\beqn
 && 2{\rm Re}\Ll Res.\Lm\Ls J_1(z_n) + J_2(z_n) \Rs
                 \frac{2\pi i}{\exp(-2\pi i z_n)-1}
                 ; z_n=il\Rm\Rl \nn\\
 &=&\frac{8\pi^3(1+e_k)(k^2-l^2-\lambda^2)}
         {e_k^2e_l l(k+l+\lambda)(k+l-\lambda)(k-l+\lambda)(k-l-\lambda)} 
         \nn \\
 &&  -\Ll k \leftrightarrow l~{\rm for~the ~first~term} \Rl
   -\frac{8\pi^4(1+e_l)(2+e_l)}
         {e_ke_l^3 kl}, \\
 && 2{\rm Re}\Ll Res.\Lm\Ls J_3(z_n) + J_4(z_n) \Rs
                 \frac{2\pi i}{\exp(-2\pi i z_n)-1}
                 ; z_n=il\Rm\Rl \nn\\
 &=&-\frac{16\pi^3(1+e_l)\lambda}
         {e_\lambda e_l^2 (k+l+\lambda)(k+l-\lambda)(k-l+\lambda)(k-l-\lambda)}
   +\frac{8\pi^4(1+e_l)(2+e_l)}
         {e_\lambda e_l^3 l\lambda} \nn \\
\eeqn
for the pole $z_n=il$, 
\beqn
 && 2{\rm Re}\Ll Res.\Lm\Ls J_1(z_n) + J_2(z_n)\Rs
                 \frac{2\pi i}{\exp(-2\pi i z_n)-1}
                 ; z_n=i(k+\lambda)\Rm\Rl \nn\\
 &=&-\frac{4\pi^3(e_k+e_{k+\lambda}+2e_k e_{k+\lambda})}
          {e_k^2e_{k+\lambda}^2 k((k+\lambda)^2-l^2)}, \\
 && 2{\rm Re}\Ll Res.\Lm\Ls J_3(z_n) + J_4(z_n)\Rs
                 \frac{2\pi i}{\exp(-2\pi i z_n)-1}
                 ; z_n=i(k+\lambda)\Rm\Rl \nn\\
 &=&-\frac{4\pi^3(1+e_{k+\lambda})}
          {e_\lambda e_{k+\lambda}^2 k((k+\lambda)^2-l^2)}
\eeqn
for the pole $z_n=i(k+\lambda)$, and  
\beqn
 && 2{\rm Re}\Ll Res.\Lm\Ls J_1(z_n) + J_2(z_n) \Rs
                 \frac{2\pi i}{\exp(-2\pi i z_n)-1}
                 ; z_n=i\abs{k-\lambda}\Rm\Rl \nn\\
 &=&-\frac{4\pi^3(1+e_k)(2e_k+e_k^2-e_{\lambda})}
          {e_k^2(e_k-e_{\lambda})^2 k((k-\lambda)^2-l^2)}
    +\frac{4\pi^3(1+e_k)}
          {e_k^2 k((k-\lambda)^2-l^2)}\theta(k-\lambda), \\
 && 2{\rm Re}\Ll Res.\Lm\Ls J_3(z_n) + J_4(z_n) \Rs
                 \frac{2\pi i}{\exp(-2\pi i z_n)-1}
                 ; z_n=i\abs{k-\lambda}\Rm\Rl \nn\\
 &=&\frac{4\pi^3(1+e_k)(1+e_\lambda)}
         {e_\lambda(e_k-e_{\lambda})^2 k((k-\lambda)^2-l^2)}
\eeqn
for the pole $z_n=i\abs{k-\lambda}$.

In summary, the contribution of this part is written as 
\beqn
 &&-\frac{16\pi^3(1+e_l)\lambda}
          {e_\lambda e_l^2 
           (k+l+\lambda)(k+l-\lambda)(k-l+\lambda)(k-l-\lambda)},
           \label{partIV1}\\
 &&-\frac{4\pi^3(e_k+e_{k+\lambda}+2e_k e_{k+\lambda})}
          {e_k^2e_{k+\lambda}^2 k((k+\lambda)^2-l^2)}
    -\frac{4\pi^3(1+e_{k+\lambda})}
          {e_\lambda e_{k+\lambda}^2 k((k+\lambda)^2-l^2)},
          \label{partIV2}\\
 &&\frac{4\pi^3(1+e_k)}
         {e_k^2e_\lambda k((k-\lambda)^2-l^2)}
    +\frac{4\pi^3(1+e_k)}
          {e_k^2 k((k-\lambda)^2-l^2)}\theta(k-\lambda).
\label{partIV}
\eeqn
Note that all terms above are finite because this part corresponds to 
the diagram where the fermion and the photon wind around $S^1$.

\end{appendix}



\end{document}